\documentclass[a4paper,11pt]{article}
\usepackage{pos}
\usepackage[percent]{overpic}
\usepackage{lineno}

\title{Detection of extended TeV emission around the Geminga pulsar with H.E.S.S.}
 \ShortTitle{Detection of Geminga with H.E.S.S.}

\author*[a,c]{A.M.W. Mitchell}
\author[b]{S. Caroff}
\author[c]{J. Hinton}
\author[d]{L. Mohrmann}

\affiliation[a]{Department of Physics, ETH\,Zurich,\\
CH-8093 Zurich, Switzerland}
\affiliation[b]{Laboratoire d'Annecy de Physique des Particules, Univ. Grenoble Alpes, \\
Univ. Savoie Mont Blanc, CNRS, LAPP, 74000 Annecy, France}
\affiliation[c]{Max-Planck-Institut f\"ur Kernphysik, \\
P.O. Box 103980, D 69029 Heidelberg, Germany}
\affiliation[d]{Friedrich-Alexander-Universit\"at Erlangen-N\"urnberg, Erlangen Centre for Astroparticle Physics, \\
Erwin-Rommel-Str. 1, D 91058 Erlangen, Germany}

\forColl{H.E.S.S.} 

\emailAdd{amitchell@phys.ethz.ch}
\emailAdd{sami.caroff@lapp.in2p3.fr}

\abstract{Highly extended gamma-ray emission around the Geminga pulsar was discovered by Milagro and verified by HAWC. Despite many observations with Imaging Atmospheric Cherenkov Telescopes (IACTs), detection of gamma-ray emission on angular scales exceeding the IACT field-of-view has proven challenging. Recent developments in analysis techniques have enabled the detection of significant emission around Geminga in archival data with H.E.S.S.. 
In 2019, further data on the Geminga region were obtained with an adapted observation strategy. Following the announcement of the detection of significant TeV emission around Geminga in archival data, in this contribution we present  the detection in an independent dataset. New analysis results will be presented, and emphasis given to the technical challenges involved in observations of highly extended gamma-ray emission with IACTs. 
}

\FullConference{37$^{\rm{th}}$ International Cosmic Ray Conference (ICRC 2021)\\
		July 12th -- 23rd, 2021\\
		Online -- Berlin, Germany}

\begin{document}

\maketitle

\section{Introduction}

The curious nature of the Geminga pulsar is reflected in its name; originating from a {\bf GA}mma-ray source in {\bf Gemini}, yet also from a Milanese dialect meaning ``it's not there'' \cite{Bignami83}. Unusually for pulsars, Geminga is radio quiet and was first detected through pulsed emission in gamma-rays. It is also one of the closest pulsars to Earth, located at a mere 250\,pc away, with a spin period of 237\,ms, a characteristic age of 342\,kyr, and a spin-down luminosity of $\dot{E} = 3.2\times 10^{34}\mathrm{ergs}^{-1}$ \cite{ATNF}.

Despite the detection of extended gamma-ray emission around Geminga by Milagro and the subsequent confirmation by HAWC, detection of the emission by Imaging Atmospheric Cherenkov Telescopes (IACTs) has remained elusive \cite{HAWCgeminga,MAGICsearch}. Given its proximity and age, the emission is considerably extended, at $\sim 5.5^\circ$ radius measured by HAWC in the energy range 8-40\,TeV \cite{HAWCgeminga}. This size is beyond that of the typical IACT field of view, making its detection particularly challenging. 

Whilst IACT facilities such as H.E.S.S. observe the Cherenkov light from extensive air showers initiated by cosmic rays and gamma-rays in our atmosphere, HAWC is a ground-based particle detector that observes the Cherenkov light in water tanks for energetic extensive air showers \cite{HESScrab,HAWCcrab}.  As such, the two facilities employ distinct but complementary techniques. The H.E.S.S. (High Energy Stereoscopic System) experiment is an array of five IACTs located in the Khomas Highlands of Namibia at $\sim1800$\,m altitude. Four telescopes are located on the corners of a square with 120\,m sides, have a mirror dish area of $108\mathrm{m}^2$, a field-of-view of $5^\circ$ and are known as CT1-4 respectively. The fifth telescope (CT5) is located at the centre of the square, has a mirror dish area of $614\mathrm{m}^2$, a field-of-view of $3.2^\circ$ and is sensitive to lower energy gamma-rays. 

Recently, several discrepancies were found between HAWC and H.E.S.S. data, especially concerning more extended gamma-ray sources, and a more concerted effort was made to compare and understand data from the two experiments in the common part of the Galactic plane \cite{HAWCHESS}.
Applying some of the lessons learnt to archival H.E.S.S. data on the Geminga pulsar, extended emission could be confirmed by the H.E.S.S. experiment as shown in \cite{tevpageminga}.

In 2019, further observations of the Geminga pulsar were made, with a particularly large offset of pointing positions around the pulsar, namely $\pm1.6^\circ$ in R.A. and Dec (compared to the more usual $\sim0.7^\circ$). Offsets this large push the limits of the system capabilities in an attempt to cover as much of the extended emission as possible, given the limited H.E.S.S. field-of-view of $5^\circ$, as the relative acceptance starts to degrade considerably for offsets beyond $\gtrsim 1^\circ$ \cite{HESScrab}. For this reason the fifth telescope (CT5), with an even smaller field-of-view of $3.2^\circ$, was not included in this analysis. 
Using this independent dataset, we confirm the detection by H.E.S.S. of extended emission around the Geminga pulsar, and present first results concerning the morphology of the emission. It should be noted, however, that even with the adapted pointing positions, the full scale of the emission continues to exceed the H.E.S.S. field of view such that the true extent of the TeV emission cannot be measured.

\section{Analysis and Results}

Given the challenges posed by such highly extended emission, standard background approaches such as the Ring background or the Reflected background are ineffective \cite{bgmethods}.
In the Ring method, the background is estimated from regions free of source emission at equal distance from the source, whereas in the Reflected method the background is estimated from regions of equal size to the test region yet reflected around the pointing position of the telescope in order to provide equal camera acceptance.  Both of these methods rely on utilising source-free regions of the sky within the field-of-view; yet with observations of the Geminga region, the source fills the field-of-view such that there are no source-free regions available. 

For this reason, two different background estimation methods were used; the so-called On-Off and field-of-view (FOV) methods. The FOV method uses a model of the acceptance to predict the expected background over the full FOV, and is the approach most similar to that used by the HAWC collaboration \cite{HAWCHESS}.
The On-Off method treats all data taken on the Geminga region as "On source" data, and estimates the background from "Off source" data which are defined by the user. In this case, extragalactic data taken on regions with no significant source (e.g. observations of dwarf spheroidal galaxies) are used.  

H.E.S.S. observation data is typically collected in ``runs'' that last $\sim28$\,min; each On run is therefore matched with a suitable Off run for the analysis. Matching criteria include: the combination of telescopes present in the run, which we require to be all four of the 12\,m telescopes; the duration of the run, which must agree to within $\sim4$\,min; and the average zenith angle of the run, which must agree to within $\sim5^\circ$. Additional corrections for remaining differences are applied during the analysis. 

To gain a handle on the systematics that affect this analysis, two independent lists of Off runs were prepared; variation in the results between the two can then be used as a measure of the intrinsic systematic uncertainties of this approach. 
All results were analysed using a sensitive image template-based analysis \cite{impact} and cross-checked using an independent analysis chain \cite{deNaurois}.  

\begin{figure}
\centering
\begin{overpic}[width=0.49\textwidth]{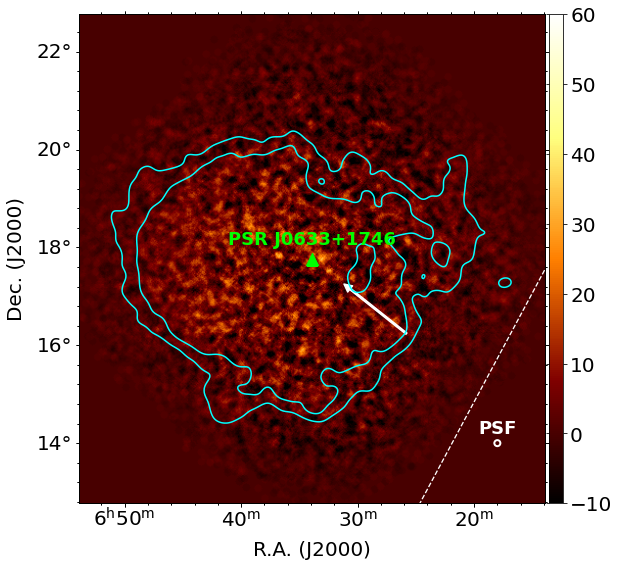}
\put(30,80){\textcolor{white}{\large{\bf PRELIMINARY}}}
\end{overpic}
\begin{overpic}[width=0.49\textwidth]{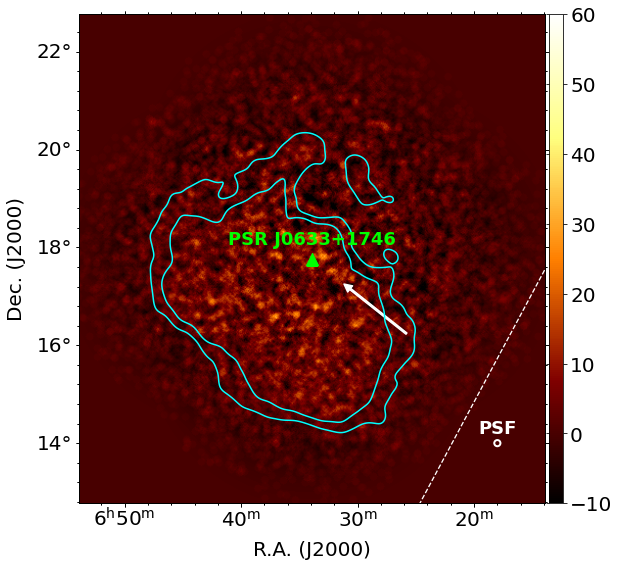}
\put(30,80){\textcolor{white}{\large{\bf PRELIMINARY}}}
\end{overpic}

\caption{Excess emission around the Geminga pulsar with two different background estimation methods. Left: using the On-Off background method and Right: using the FOV background method. }
\label{fig:maps}
\end{figure}

Figure \ref{fig:maps} shows excess maps of the Geminga region, with a correlation radius of $0.08^\circ$, slightly larger than the point spread function (PSF) in both analyses. Overlaid on the maps are the pulsar location, the proper motion direction of the pulsar (white arrow) \cite{HobbsPM} and the Galactic plane (white dashed line). Cyan contours are obtained with a $0.5^\circ$ correlation radius and additionally smoothed, whilst the PSF is indicated on the maps for scale. 
There are three things to note in figure \ref{fig:maps}: firstly, that the emission is not centred on the pulsar location but appears offset; secondly, that the emission distribution is not identical between the two background methods; and thirdly, that there is no clear distinction / edge to the emission. 
We address each of these points in turn below. 

\subsection{Offset and asymmetric emission}

\begin{figure}
\centering
\begin{overpic}[width=0.6\textwidth]{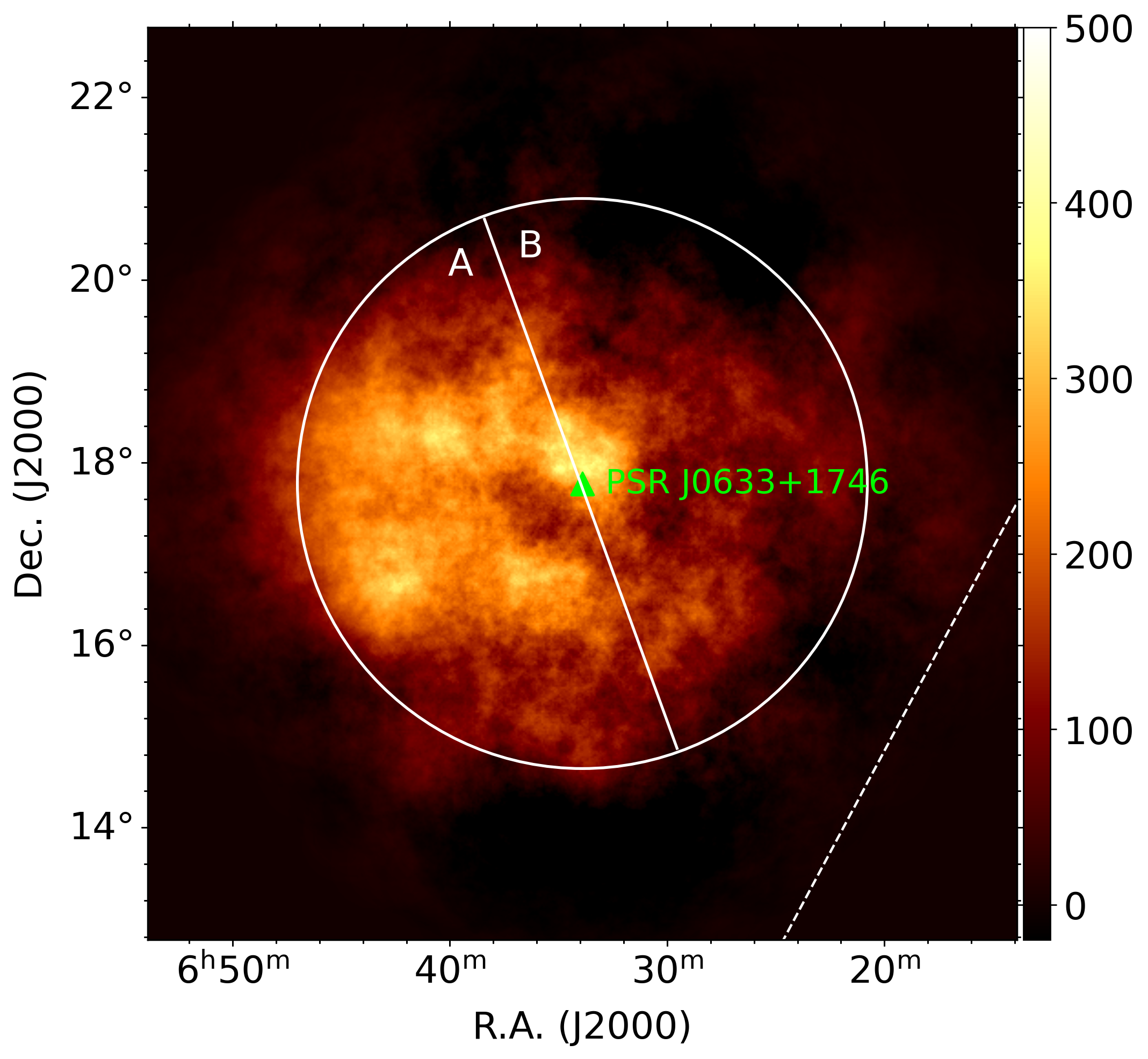}
\put(30,80){\textcolor{white}{\large{\bf PRELIMINARY}}}
\end{overpic}
\caption{Excess emission in the Geminga region, correlated with $0.5^\circ$ radius. To establish whether the emission is isotropic, two regions (A \& B) were defined as described in the main text. There is a clear asymmetry to the emission, with side A a factor $\sim1.5$ brighter than side B. }
\label{fig:split}
\end{figure}

To verify the apparent offset and asymmetric nature to the TeV emission, we first investigated the azimuthal distribution of emission around the pulsar (within a $3^\circ$ radius), finding the direction in which the emission was seen to peak and averaging this over different background methods. Defining a line perpendicular to this, we split the region into two as shown in figure \ref{fig:split}. Radial profiles of the emission were constructed out to $3^\circ$ radius (as shown by the circle in figure \ref{fig:split}) and the ratio of the profiles from sides A and B were used to determine if there was a significant difference in brightness. We find that side A is indeed systematically brighter than side B by a factor $\sim1.5$, confirming this apparent asymmetry to the emission.

From figure \ref{fig:split}, one might be forgiven for thinking that there is additional substructure to the emission around Geminga; however, we conducted several tests varying the excess counts within statistical uncertainties, concluding that this sub-structure is not statistically significant. The discrepancy between background methods illustrated in figure \ref{fig:maps} supports this assessment. 

\subsection{Discrepancy in morphology between background methods}

\begin{figure}
\centering
\begin{overpic}[width=0.49\textwidth]{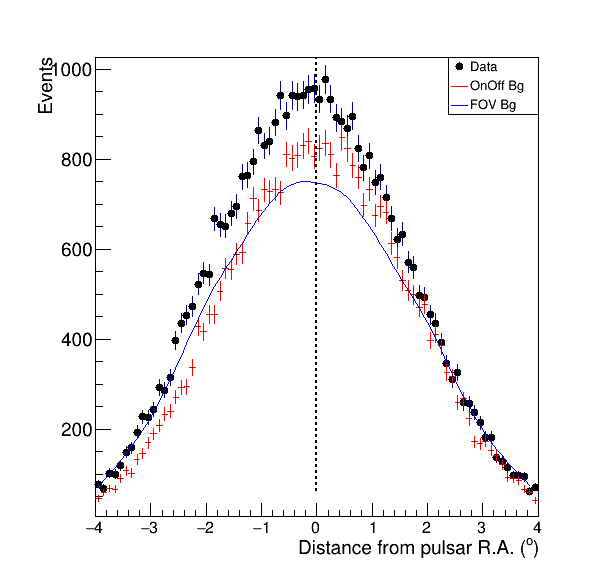}
\put(30,30){\textcolor{red}{\large{\bf PRELIMINARY}}}
\end{overpic}
\begin{overpic}[width=0.49\textwidth]{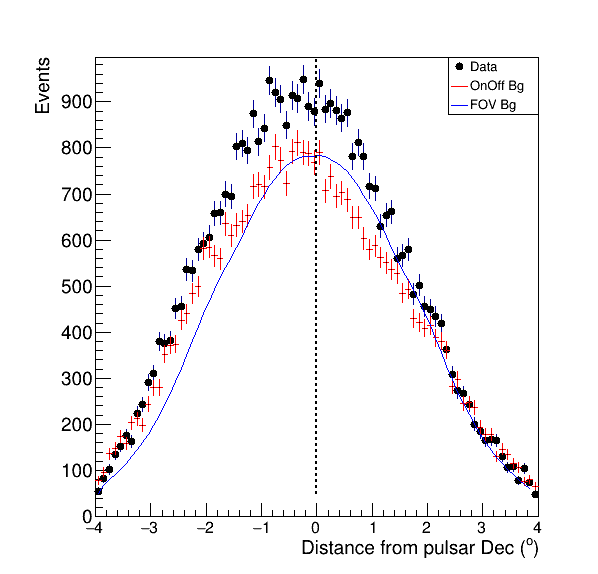}
\put(30,30){\textcolor{red}{\large{\bf PRELIMINARY}}}
\end{overpic}

\caption{Projections of the excess emission onto the R.A. (left) and Dec (right) axes, compared to background using two different methods. A clear excess is seen over much of the field of view, with some discrepancies between the two background methods. }
\label{fig:slices}
\end{figure}

In order to directly compare the two background methods, the uncorrelated On counts were projected onto the R.A. and Dec axes, together with the background counts from the two methods. These slice projections are shown in figure \ref{fig:slices}, where the differences between levels of background counts agrees with the morphology shown in figure \ref{fig:maps}. For example, towards lower declinations than the pulsar position, the FOV background is lower than the On-Off background, such that the emission would consequently be more pronounced in lower declinations for FOV background -- as indeed shown in figure \ref{fig:maps}. 

It should be noted that the FOV background is normalised outside of an exclusion region, defined to be a circle around Geminga with a $3.2^\circ$ radius, whereas this is not the case for the On-Off method. Additionally, the On-Off analysis was found to be very sensitive to the choice of Off data, despite the corrections applied at analysis stage. It is therefore important that Off data is chosen carefully to also be comparable in terms of atmospheric conditions and the status of the instrument. Ideally, dedicated Off data would be taken specifically for such an analysis, although observation time constraints typically do not allow for such an intensive programme. 

\subsection{Radial extent of the emission}

\begin{figure}
\centering
\begin{overpic}[width=0.7\textwidth]{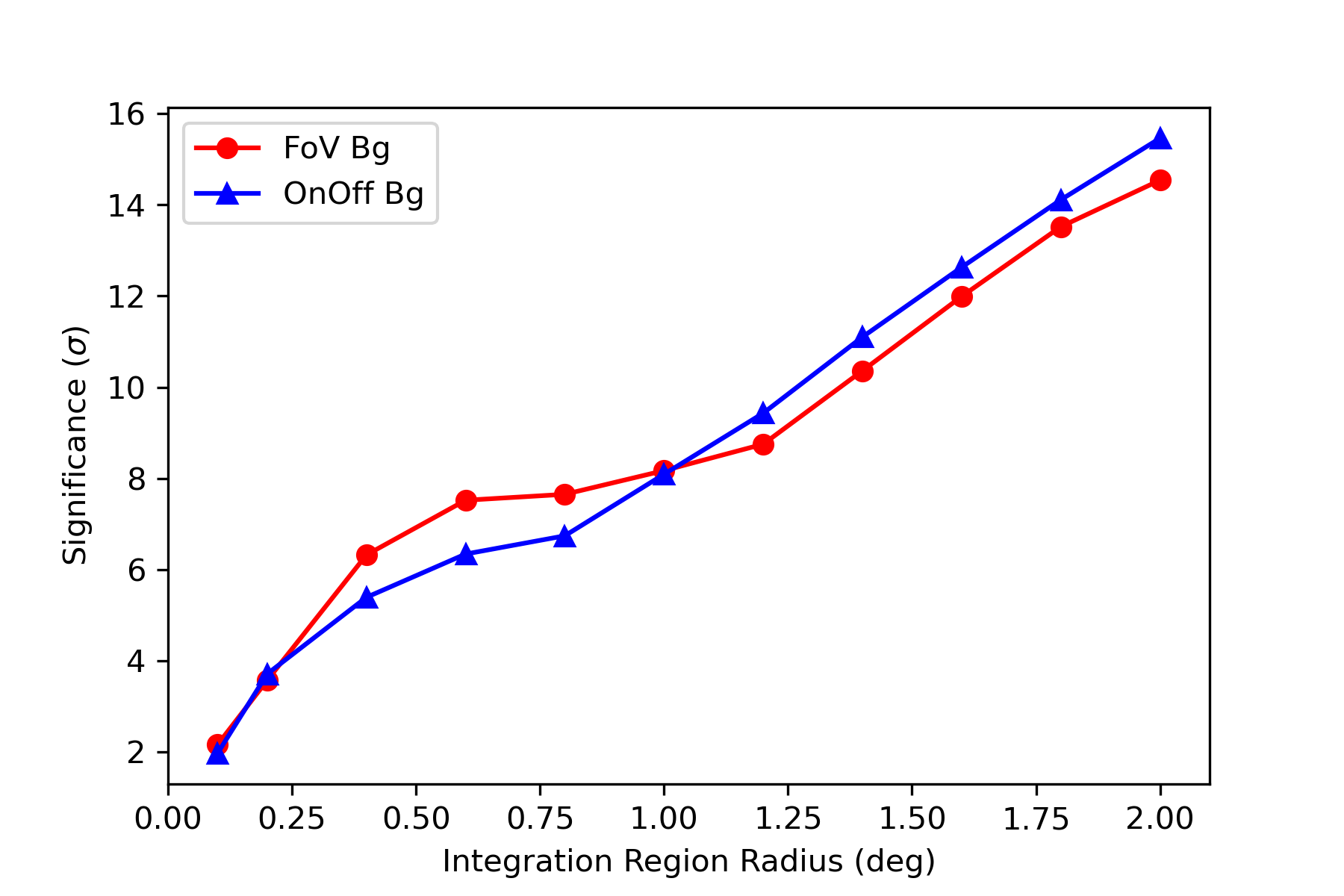}
\put(35,50){\textcolor{red}{\large{\bf PRELIMINARY}}}
\end{overpic}
\caption{The significance is seen to increase with radius of the integration region out to $2^\circ$ without noticeable saturation. A radius of $2^\circ$ is therefore a lower limit on the true extent of the emission. }
\label{fig:radsigma}
\end{figure}

The integration region is the region treated as ``On source'' from which the significance is estimated. Figure \ref{fig:radsigma} shows how the significance of the gamma-ray emission increases with increasing radius of the integration region without flattening out to $2^\circ$. This lack of flattening to the curve allows us to confirm that the gamma-ray emission extends beyond $2^\circ$, yet does not enable the true extent of the emission to be measured. 
Integration regions larger than this are challenging to analyse, such that we merely state here that we expect that gamma-ray emission from around the Geminga pulsar continues to fill the field-of-view of H.E.S.S., despite our efforts with a much increased pointing offset of $1.6^\circ$ in this dataset. 

\section{Conclusions and Outlook}

H.E.S.S. confirms the detection of extended gamma-ray emission around the Geminga pulsar in an independent dataset from that of the previously announced detection. Analysis of such TeV sources with IACTs does, however, remain challenging. As particle detectors such as HAWC and LHAASO discover increasingly more extended TeV sources around middle-aged pulsars; many of them likely to be in the halo evolutionary phase; detailed studies of such sources with the IACT capabilities (improved energy and angular resolution) are of great interest. 

In this contribution we describe some of the challenges IACTs are faced with in the analysis of Geminga, and some of the steps we have taken to check and verify our results internally. Systematic uncertainties of such a measurement are highly relevant to the interpretation of the results; the distribution of the emission is found to be asymmetric around the pulsar, but we do not claim further specific details of sub-structure to the morphology. Such features are below the sensitivity that our analysis can confirm within the level of the uncertainties. 

Nevertheless, we can state that the true radial extent of the emission is definitely $\geq 2^\circ$ and highly likely to be $\geq 3.2^\circ$. 
A forthcoming publication will describe this analysis in detail, including the tests made with different background approaches. This will additionally include a spectral analysis of the innermost region around the Geminga pulsar, and a radial profile of the emission, to which a diffusion model will be applied. 
We look forward to releasing these results to the community in due course.

\subsection*{Acknowledgements}

\noindent \small{The support of the Namibian authorities and of the University of Namibia in facilitating the construction and operation of H.E.S.S. is gratefully acknowledged, as is the support by the German Ministry for Education and Research (BMBF), the Max Planck Society, the German Research Foundation (DFG), the Helmholtz Association, the Alexander von Humboldt Foundation, the French Ministry of Higher Education, Research and Innovation, the Centre National de la Recherche Scientifique (CNRS/IN2P3 and CNRS/INSU), the Commissariat \`{a} l'\'{e}nergie atomique et aux \'{e}nergies alternatives (CEA), the U.K. Science and Technology Facilities Council (STFC), the Knut and Alice Wallenberg Foundation, the National Science Centre, Poland grant no. 2016/22/M/ST9/00382, the South African Department of Science and Technology and National Research Foundation, the University of Namibia, the National Commission on Research, Science \& Technology of Namibia (NCRST), the Austrian Federal Ministry of Education, Science and Research and the Austrian Science Fund (FWF), the Australian Research Council (ARC), the Japan Society for the Promotion of Science and by the University of Amsterdam.

We appreciate the excellent work of the technical support staff in Berlin, Zeuthen, Heidelberg, Palaiseau, Paris, Saclay, T\"{u}bingen and in Namibia in the construction and operation of the equipment. This work benefitted from services provided by the H.E.S.S. Virtual Organisation, supported by the national resource providers of the EGI Federation.}

\section*{Full Authors List: \Coll\ Collaboration}
%
%
\scriptsize
\noindent
H.~Abdalla$^{1}$, 
F.~Aharonian$^{2,3,4}$, 
F.~Ait~Benkhali$^{3}$, 
E.O.~Ang\"uner$^{5}$, 
C.~Arcaro$^{6}$, 
C.~Armand$^{7}$, 
T.~Armstrong$^{8}$, 
H.~Ashkar$^{9}$, 
M.~Backes$^{1,6}$, 
V.~Baghmanyan$^{10}$
% ...

\medskip

\noindent
$^{1}$University of Namibia, Department of Physics, Private Bag 13301, Windhoek 10005, Namibia\\
$^{2}$Dublin Institute for Advanced Studies, 31 Fitzwilliam Place, Dublin 2, Ireland\\
$^{3}$Max-Planck-Institut f\"ur Kernphysik, P.O. Box 103980, D 69029 Heidelberg, Germany\\
$^{4}$High Energy Astrophysics Laboratory, RAU,  123 Hovsep Emin St  Yerevan 0051, Armenia\\
$^{5}$Aix Marseille Universit\'e, CNRS/IN2P3, CPPM, Marseille, France\\
$^{6}$Centre for Space Research, North-West University, Potchefstroom 2520, South Africa\\
$^{7}$Laboratoire d'Annecy de Physique des Particules, Univ. Grenoble Alpes, Univ. Savoie Mont Blanc, CNRS, LAPP, 74000 Annecy, France\\
$^{8}$University of Oxford, Department of Physics, Denys Wilkinson Building, Keble Road, Oxford OX1 3RH, UK\\
$^{9}$IRFU, CEA, Universit\'e Paris-Saclay, F-91191 Gif-sur-Yvette, France\\
$^{10}$Instytut Fizyki J\c{a}drowej PAN, ul. Radzikowskiego 152, 31-342 Krak{\'o}w, Poland\\
% ...

\end{document}